\documentclass[twocolumn,showpacs]{revtex4}

\usepackage{graphicx}
\graphicspath{{pict/}{}}

\usepackage{color}

\begin{document}

\title{Discrete surface solitons in two-dimensional anisotropic photonic lattices}

\author{Rodrigo A. Vicencio and Sergej Flach}

\affiliation{Max-Planck-Institut f\"ur Physik Komplexer Systeme, N\"othnitzer Str. 38, Dresden 01187, Germany}

\author{Mario I. Molina}

\affiliation{Departmento de F\'{\i}sica, Facultad de Ciencias,
Universidad de Chile, Casilla 653, Santiago, Chile}

\author{Yuri S. Kivshar}

\affiliation{Nonlinear Physics Center, Research School of Physical
Sciences and Engineering, Australian National University, Canberra
ACT 0200, Australia}

\begin{abstract}
We study nonlinear surface modes in two-dimensional {\em
anisotropic} periodic photonic lattices and demonstrate that, in a
sharp contrast to one-dimensional discrete surface solitons, the
mode threshold power is lower at the surface, and two-dimensional
discrete solitons can be generated easier near the lattice corners
and edges. We analyze the crossover between effectively one- and
two-dimensional regimes of the surface-mediated beam localization in
the lattice.
\end{abstract}

%\ocis{030.1640, 190.4420}

\maketitle

Surface modes have been studied in different branches of physics; in
guided wave optics surface states were predicted to exist at
interfaces separating periodic and homogeneous dielectric
media~\cite{optics}. The interest in studying surface waves has
been renewed recently because the interplay of discreteness and nonlinearity can facilitate the formation of discrete surface
solitons~\cite{makris,suntsov} at the edge of the waveguide array. That can be understood as the localization of a discrete optical
soliton near the surface~\cite{surfol} for powers exceeding a
certain threshold value, for which the repulsive effect of the surface is balanced. A similar effect of light localization near the edge of the waveguide array and the formation of surface gap solitons have been predicted and observed for
defocusing nonlinear media~\cite{PRL_kartashov,PRL_canberra}.

It is important to analyze how the properties of nonlinear surface
waves are modified by the lattice dimensionality, and the first
studies of different types of discrete surface solitons in
two-dimensional lattices~\cite{ol_kar1,ol_kar2,ol_moti,pre_arxiv}
revealed, in particular, that the presence of a surface increases
the stability region for two-dimensional (2D) discrete
solitons~\cite{pre_arxiv} and the threshold power for the edge
surface state is slightly higher than that for the corner
soliton~\cite{ol_moti}.

In this Letter we consider {\em anisotropic} semi-infinite
two-dimensional photonic lattices and study the crossover between
one- and two-dimensional surface solitons emphasizing
the crucial effect of the lattice dimensionality on the formation of
surface solitons.

%
%%%%%%%%%%%%%%%%%%%%Fig1%%%%%%%%%%%%%%%%%%%%%%%%%%%%%%%%%
\begin{figure}[htbp]
\centerline{\scalebox{0.52}{\includegraphics{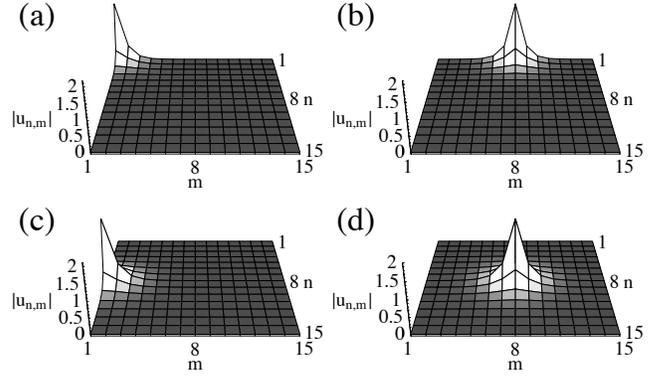}}}
\caption{Examples of (a) corner, (b,c) two edge surface, and (d)
central localized nonlinear modes for $\lambda=5$.} \label{fig1}
\end{figure}
%%%%%%%%%%%%%%%%%%%%%%%%%%%%%%%%%%%%%%%%%%%%%%%%%%%%%%%%%%
%

\begin{figure}[htbp]
\centerline{\scalebox{0.48}{\includegraphics{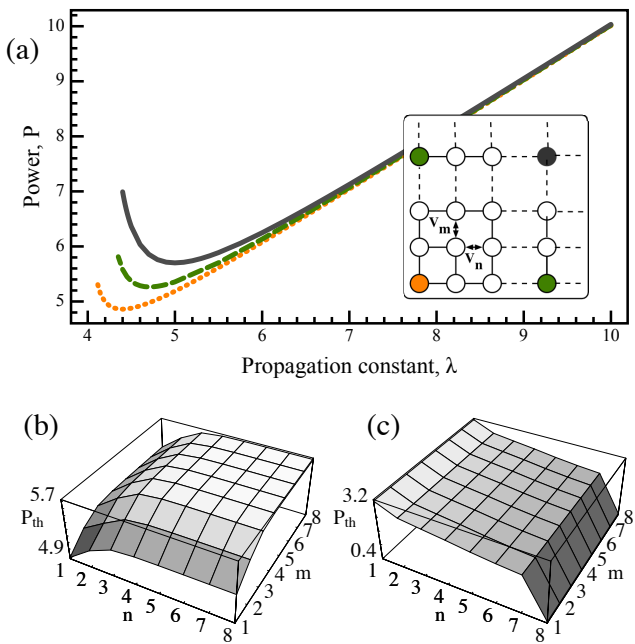}}}
\caption{(Color online) (a) Power diagram of the central
(continuous), edges surface (dashed), and corner (dotted) localized
modes for the isotropic case ($\alpha=1$). Inset: Schematic of a
semi-infinite two-dimensional lattice. (b,c) Examples of the
threshold power landscapes $P_{\rm th}(n,m)$ for $\alpha=1$ and
$\alpha=0$, respectively.}\label{fig2}
\end{figure}

We consider a semi-infinite 2D lattice [shown schematically in
Fig.\ref{fig2}(a) below], described by the system of coupled-mode
equations for the normalized amplitudes
$u_{n,m}$~\cite{led2d,yoMag},
\begin{equation}
i \frac{\partial u_{n,m}}{\partial \xi} + \left(V_n+V_m\right) u_{n,m}
+ |u_{n,m}|^{2} u_{n,m} =0, \label{eq}
\end{equation}
where $\xi$ is the normalized propagation distance. We define the
lattice coupling as follows:
\begin{eqnarray*}
V_n u_{n,m}=\left\{\begin{array}{c} u_{2,m}\hspace{1.5cm} n=1, m\geq 1, \\
u_{n+1,m}+u_{n-1,m}\hspace{0.5cm} n >1, \end{array} \right.\\
V_m u_{n,m}=\alpha \left\{\begin{array}{c} u_{n,2} \hspace{2cm} n
\geq 1, m=1, \\  (u_{n,m+1}+u_{n,m-1}) \hspace{0.5cm} m >1,
\end{array}\right.
\end{eqnarray*}
where  $\alpha$ characterizes the lattice anisotropy.

Linear lattice waves of the form $u_{n,m}(\xi)=u_0 \sin(kn) \sin(qm)
\exp(i\beta \xi)$ satisfy the dispersion relation $\beta_{kq}=2(\cos
k+\alpha \cos q)$. In the nonlinear case, we look for localized
stationary solutions of the form $u_{n,m}(\xi)=u_{n,m} \exp(i \lambda
\xi)$, where the amplitudes $u_{n,m}$  are real, and $\lambda$ is the
nonlinear propagation constant. For a given $\lambda$, localized
solutions are found in a $15\times 15$ lattice by using the Newton-Raphson method.

We calculate the power threshold $P_{\rm th}$ that characterizes the
discrete solitons in 2D lattices~\cite{flachthres}. We study three
different modes: corner [Fig.~\ref{fig1}(a)], edge surface
[Fig.~\ref{fig1}(b,c)] and central [Fig.~\ref{fig1}(d)] localized
modes. The corner and edge modes represent 2D surface localized
modes, and the central mode corresponds to a 2D discrete soliton of
an infinite lattice. We find the total power $P = \sum_{n,m}
|u_{n,m}|^{2}$ of all those modes and perform a linear stability
analysis~\cite{yoMag,surfol} for each solution [Fig.~\ref{fig2}(a)].

We observe that the threshold power~\text{$P_{\rm th}$} for the
surface modes is smaller than the power corresponding to the central
mode, with the corner state having the smallest ~\text{$P_{\rm
th}$}. This interesting feature was previously observed for a
single nonlinear impurity placed near a boundary of a 2D
lattice~\cite{mm_prb}, and it appears also in the anisotropic
model. Therefore, in a sharp contrast with one-dimensional (1D)
surface solitons, the surface of a 2D lattice creates an effectively
attractive potential for the localized modes that reduces the
threshold power for the mode localization.

%
%%%%%%%%%%%%%%%%%%%%  Fig2  %%%%%%%%%%%%%%%%%%%%%%%%%%%%%%%%%%%%%%%%%%%%%%%%%%%
%%%%%%%%%%%%%%%%%%%%%%%%%%%%%%%%%%%%%%%%%%%%%%%%%%%%%%%%%%%%%%%%%%%%%%%%%%%%%%

The linear stability analysis of the surface localized modes
coincides with the Vakhitov-Kolokolov stability
criterion~\cite{book}: the surface modes are stable for
$dP/d\lambda>0$, and unstable otherwise. To study in more details
how the threshold power $P_{\rm th}$ varies for the modes localized
at different points of the lattice, we compute \textit{the power
threshold landscape}: We look numerically for one-peak localized
modes centered at different sites of one-quarter of the whole
lattice and find the first stable solution at the threshold power
which corresponds to $P_{\rm th}$ at the site $(n,m)$. In
Fig.~\ref{fig2}(b,c), we show the examples of two landscapes for two
different values of the lattice anisotropy $\alpha$. For $\alpha=1$,
$P_{\rm th}$ grows from the minimum $P_{\rm th} \approx4.9$ at the corner
$(1,1)$ to two maxima corresponding to the edge
surface modes $(8,1)$ and $(1,8)$ and the central mode $(8,8)$. For
$\alpha=0$, we observe that the required power to excite a localized
mode decreases from the surface ($n=1$) to the center ($n=8$). The
value $P_{th}(1,m) \approx 3.2$ corresponds to the threshold power
of discrete surface solitons in a semi-infinite array~\cite{surfol}.
For 1D localized modes there exists no power threshold in the
continuum limit~\cite{flachthres}, but in our system $P_{\rm
th}(8,m)\approx 0.4$  due to finite-size effects.

\begin{figure}[htbp]\centerline{\scalebox{0.46}{\includegraphics{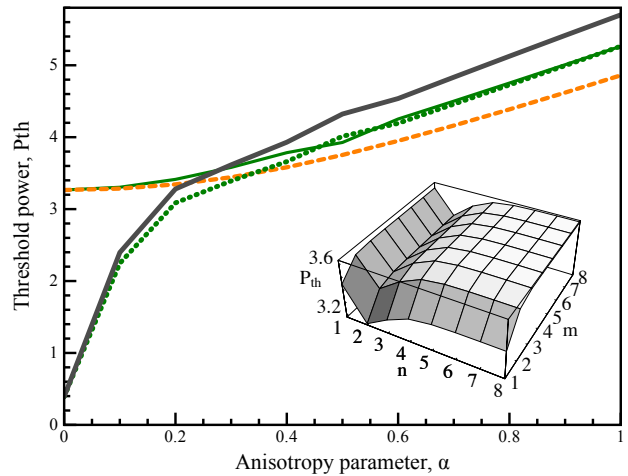}}}
\caption{(Color online) Threshold power $P_{\rm th}$ vs. $\alpha$
for the central (continuous thick), two edge surface (continuous
thin and dotted), and corner (dashed line) localized modes. Inset:
$P_{\rm th}(n,m)$ for $\alpha=0.3$.} \label{fig3}
\end{figure}

Next, we study the effect of the lattice anisotropy on the power
threshold of discrete surface solitons. In Fig.~\ref{fig3} we show
$P_{\rm th}$ for different values of the parameter $\alpha$
($0\leq\alpha\leq1$), for the four types of localized modes shown in
Figs.~\ref{fig1}(a-d).

%%%%%%%%%%%%%%%%%%%%%Fig3%%%%%%%%%%%%%%%%%%%%%%%%%%%%%%%%%%%%%
%%%%%%%%%%%%%%%%%%%%%%%%%%%%%%%%%%%%%%%%%%%%%%%%%%%%%%%%%%%%%%%

When the anisotropy parameter $\alpha$ grows, the threshold power
grows for all localized modes. It means that, by increasing the
effective dimensionality of the system, the required power to excite
a nonlinear discrete localized state also increases. The threshold
power for two edge surface modes decreases as $\alpha$ decreases to
$\alpha \approx 0.3$, when the power curves cross and diverge. For
$\alpha=0$, the edge soliton centered at $(8,1)$
[Fig.~\ref{fig1}(c), dotted line in Fig.~\ref{fig3}] only interacts
with an effective 1D array ($m=1$). On the contrary, $P_{\rm th}$
for the edge surface mode centered at $(1,8)$ [Fig.~\ref{fig1}(b),
continuous-thin line in Fig.~\ref{fig3}] approaches the same limit
as the corner mode. For $\alpha=0$, this mode only interacts with an
effective 1D chain ($m=8$) and the surface. An interesting feature
of Fig.~\ref{fig3} is that it shows the existence of a critical
value of the lattice anisotropy, $\alpha \approx 0.3$ where all
powers thresholds almost coincide. In other words, there exists a
critical value of the lattice anisotropy where we observe a
crossover between 1D and 2D lattices.

To study this interesting effect in more details, we calculate the
power threshold landscape $P_{th}$ for $\alpha=0.3$ [see inset in
Fig.~\ref{fig3}], and observe that the localized mode centered at
$(2,m)$ possesses the lowest value of $P_{th}$ in the array and, in
particular, the localized mode centered at $(2,1)$ is the mode with
the lowest threshold power. For $\alpha\lesssim 0.3$, the minimum $P_{\rm th}$ will be located outside of the array boundaries, and {\em intermediate} localized states (located between the center and the surfaces of the array) will be the easier states to excite. For 1D semi-infinite
arrays~\cite{surfol}, it was shown that the power threshold for
surface modes decreases as the mode moves away from the surface [see
also Fig.~\ref{fig2}(c)].

%
%%%%%%%%%%%%%%%%%%%% Fig4 %%%%%%%%%%%%%%%%%%%
\begin{figure}[htbp]
\centerline{\scalebox{0.57}{\includegraphics{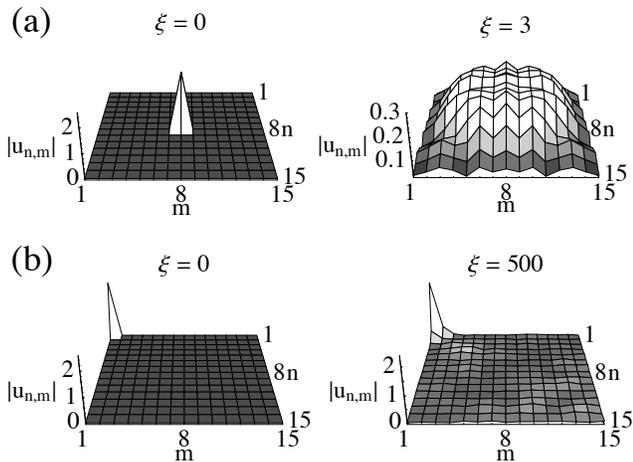}}}
\caption{Generation of surface modes by one-site excitation
($\alpha=1$) at (a) center and  (b) corner of the lattice.}
\label{fig4}
\end{figure}
%%%%%%%%%%%%%%%%%%%%%%%%%%%%%%%%%%%%%%%%%%%%%%%

Finally, we study the dynamic generation of localized modes in
two-dimensional lattices. We simulate numerically Eqs.~(\ref{eq}) by
using a one-site excitation $u_0\
\delta_{n,n'}\delta_{m,m'}$. In Fig.~\ref{fig4}, we show our results
by plotting the evolution of the mode amplitude $|u_{n,m}|$ for two
different values of $\xi$. As an example, we show the excitation of
the corner and central localized modes for the initial amplitude
$u_0=2.5$. If we try to excite a localized mode at the middle of the
lattice for this value of $u_0$, this localized state decays
quickly, as shown in Fig.~\ref{fig4}(a), and the beam diffracts
because the input power is below the effective dynamical power
threshold. On the contrary, by using the same input power we are
able to excite the corner soliton [see Fig.~\ref{fig4}(b)] which is
dynamically stable in the propagation. It is worth mentioning that,
in the one-dimensional limit ($\alpha=0$) the situation becomes {\em
completely different}: If we try to excite a surface mode at $n=1$
with the power smaller than a threshold value, but still large enough, the input beam will
not decay but will instead move to the neighboring lattice site
creating a localized mode there~\cite{makris}.

In conclusion, we have analyzed the properties of discrete surface
solitons in two-dimensional anisotropic photonic lattices, and
studied the crossover between the one- and
two-dimensional regimes of the surface-mediated beam localization in
the lattice. In particular, unlike one-dimensional discrete surface
solitons, the threshold power of the two-dimensional discrete
soliton is lowered by the surface, so that two-dimensional solitons
can be generated easier near the lattice corners and edges.

This work was supported by Fondecyt  grants 1050193 and 7050173
and the Australian Research Council.


\begin{thebibliography}{99}

\bibitem{optics} P. Yeh, A. Yariv, and A.Y. Cho, \apl {\bf 32},
102 (1978).

\bibitem{makris} K.G. Makris, S. Suntsov, D.N.
Christodoulides, G.I. Stegeman, and A. Hache, \ol {\bf
30}, 2466 (2005).

\bibitem{suntsov} S. Suntsov, K.G. Makris, D.N.
Christodoulides, G.I. Stegeman, A. Hach¶e, R. Morandotti, H. Yang,
G. Salamo, and M. Sorel, \prl {\bf 96}, 063901
(2006).

\bibitem{surfol} M.I. Molina, R.A. Vicencio, and Yu.S
Kivshar, \ol {\bf 31}, 1693 (2006).

\bibitem{PRL_kartashov}
Ya.V. Kartashov, V.A. Vysloukh, and L. Torner, \prl {\bf 96}, 073901 (2006).

\bibitem{PRL_canberra} C.R.
Rosberg, D.N. Neshev, W. Krolikowski, A. Mitchell, R.A. Vicencio,
M.I. Molina, and Yu.S. Kivshar, \prl {\bf 97}, 083901
(2006).

\bibitem{ol_kar1} Y.V. Kartashov and L. Torner, \ol {\bf 31},
2172 (2006)

\bibitem{ol_kar2} Y.V. Kartashov, V.A. Vysloukh, D. Mihalache, and L. Torner, \ol {\bf 31}, 2329 (2006).

\bibitem{ol_moti} K.G. Makris, J. Hudock, D.N. Christodoulides, G.I. Stegeman, O. Manela, and M. Segev, \ol {\bf 31}, 2774 (2006).

\bibitem{pre_arxiv} H. Susanto, P.G. Kevrekidis, B.A.
Malomed, R. Carretero-Gonzalez, and D.J. Franzeskakis,
ArXiv:nlin.PS/0607063 (2006).

\bibitem{led2d} T. Pertsch, U. Peschel, F.
Lederer, J. Burghoff, M. Will, S. Nolte, and A. T\"unnermann, \ol
{\bf 29}, 468 (2004).

\bibitem{yoMag} R.A. Vicencio and M.
Johansson, \pre {\bf 73}, 046602 (2006).

\bibitem{flachthres} S.
Flach, K. Kladko, and R.S. MacKay, \prl {\bf 78}, 1207
(1997).

\bibitem{mm_prb} M.I. Molina, \prb {\bf 74}, 045412 (2006).

\bibitem{book}
Yu.S. Kivshar and G.P. Agrawal, {\em {Optical Solitons: From
Fibers to Photonic Crystals}} (Academic, San Diego,
2003).

\end{thebibliography}
\end{document}